\documentclass[10pt,letterpaper,twocolumn]{article}
\usepackage{ol2}
\usepackage{amsmath}

\begin{document}

\twocolumn[
\title{Electromagnetic shock wave in nonlinear vacuum: Exact solution}

\author{Lubomir M. Kovachev,$^{1,*}$ Daniela A. Georgieva,$^2$ and Kamen L. Kovachev$^{1}$}

\address{$^1$Institute of Electronics, Bulgarian Academy of Sciences,\\
Tzarigradcko shossee 72,1784 Sofia, Bulgaria,
\\
$^2$Faculty of Applied Mathematics and Computer Science,\\ Technical
University of Sofia, 8, Kliment Ohridski Blvd., 1000
Sofia, Bulgaria \\
$^*$Corresponding author: lubomirkovach@yahoo.com}

\begin{abstract}
An analytical approach to the theory of electromagnetic waves in
nonlinear vacuum is developed. The evolution of the pulse is
governed by a system of nonlinear wave vector equations. Exact
solution with its own angular momentum  in form of a shock wave is
obtained.
\end{abstract}
\ocis{190.5940, 260.5950.}
]

Contemporary hight-power laser facilities can generate optical
pulses with intensities of the order of $10^{22}$ $W/cm^2$.  At the
same time the critical power for observation self-action effects due
to virtual electron-positron pairs is of order \cite{BOYD,BUL1,
BUL2} $ P_{cr}=\lambda^2/8n_0n_2=2.5-4.4\times 10^{24}\; W,$ at a
wavelength $1$ $\mu m$. Thus, for a laser pulse with waist
$r_{\bot}=1$ $mm$ the corresponding intensity becomes
$I_{cr}^{vac}=P_{cr}/r_{\bot}^2=2.5-4.4\times 10^{26} \; W/cm^2 $,
which is  above the range of the new high-power lasers. The
nonlinear addition to the refractive index in vacuum depends also on
the magnetic field. That is why new different nonlinear effects can
be expected. There are not only self-action effects, but also vacuum
birefringence \cite{KAP, ROZ1}, different kinds of four wave
interaction \cite{KL,LUN,TOM} and higher order harmonic generation
\cite{FED}. In this paper we shall investigate the self-action
effect only for intensities of the order of $I_{cr}^{vac}$.

Euler,  Heisenberg  and Kockel \cite{EULER1, EULER2} predicted
intrinsic nonlinearity of the electromagnetic vacuum due to the
electron-positron nonlinear polarization. The classical
field-dependent nonlinear vacuum dielectric tensor can be written in
the form

\begin{eqnarray}
\label{EVAC} \epsilon_{ik}=\delta_{ik}+\frac{7e^4\hbar}{45\pi
m^4c^7}\left[2\left(|\vec{E}|^2-|\vec{B}|^2\right)\delta_{ik}+7B_iB_k\right],
\end{eqnarray}
where a complex form of presenting of the electrical $E_i$ and
magnetic $B_i$ components is used. Note that the term containing
$B_iB_k$ vanishes, when a localized electromagnetic wave with only
one magnetic component $B_l$ is investigated. The dielectric
response relevant to such optical pulse is thus
\begin{eqnarray}
\label{EVAC1} \epsilon_{ik}=\delta_{ik}+\frac{14e^4\hbar}{45\pi
m^4c^7}\left(|\vec{E}|^2-|\vec{B}|^2\right)\delta_{ik}.
\end{eqnarray}
In the case when the spectral width of a pulse $\triangle k_z$
exceeds the values of the main wave-vector, i.e. $\triangle
k_z\simeq k_0$, the system of amplitude equations can be reduced to
wave type \cite{KOVACH1} and in nonlinear vacuum becomes

\begin{eqnarray}
\label{EBwave} \Delta \vec{E} -\frac{1}{c^2}\frac{\partial^2
\vec{E}}{\partial t^2}+\gamma
(|\vec{E}|^2-|\vec{B}|^2)|\vec{E}=0\nonumber \\
\Delta \vec{B} -\frac{1}{c^2}\frac{\partial^2 \vec{B}}{\partial
t^2}+\gamma (|\vec{E}|^2-|\vec{B}|^2)|\vec{B}=0,
\end{eqnarray}
where $ \gamma=\frac{7k_0^2e^4\hbar}{90\pi m^4c^7}$ and $\vec{E}$,
$\vec{B}$ are the amplitude functions. Initially, we can write the
components of the electrical and magnetic fields as a vector sum of
circular and linear components $E_z;\;  E_c=iE_x-E_y; \;B_l=-B_z$.
Thus (\ref{EBwave}) is transformed in the following scalar system of
equations

\begin{eqnarray}
\label{syst} \Delta E_z -\frac{1}{c^2}\frac{\partial^2 E_z}{\partial
t^2}+\gamma(|E_z|^2+|E_c|^2-|B_l|^2)E_z=0\nonumber\\
\Delta E_c -\frac{1}{c^2}\frac{\partial^2 E_c}{\partial
t^2}+\gamma(|E_z|^2+|E_c|^2-|B_l|^2)E_c=0 \\
\Delta B_l-\frac{1}{c^2}\frac{\partial^2 B_l}{\partial
t^2}+\gamma(|E_z|^2+|E_c|^2-|B_l|^2)B_l=0.\nonumber
\end{eqnarray}

\begin{figure*}[htb]
\centerline{
\includegraphics[width=120 mm]{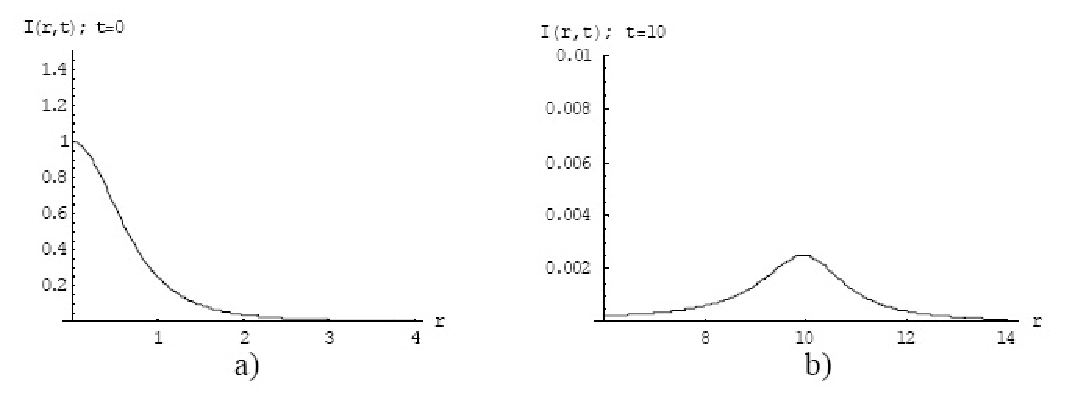}}
\caption{Time evolution of the intensity profile $I$ of the
spherically symmetric analytical solution (\ref{ANZ4}) of the linear
wave equation (\ref{W}) ($r_0=1$ and $c=1$). The initially ($t=0$)
localized amplitude function (Fig. 1a) decreases with the generation
of outside and inside fronts (Fig. 1b), while the energy density
distributes over the whole space for a finite time ($t=10$).}
\end{figure*}

Let us now parameterize the $3D+1$ space-time through
pseudospherical coordinates ($r,\tau,\theta,\varphi$):

$z=r\cosh(\tau)\cos(\theta)$,
$y=r\cosh(\tau)\sin(\theta)\sin(\varphi)$,
$x=r\cosh(\tau)\sin(\theta)\cos(\varphi)$ and $ict=r\sinh(\tau)$,
where $r=\sqrt{x^2+y^2+z^2-c^2t^2}$. After calculations the
corresponding d'Alambert operator in pseudospherical coordinates
becomes \cite{EWA}

\begin{eqnarray}
\label{dAL} \triangle-\frac{1}{c^2}\frac{\partial^2}{\partial
t^2}=\frac{3}{r}\frac{\partial}{\partial
r}+\frac{\partial^2}{\partial r^2}-\nonumber\\
\frac{1}{r^2}\frac{\partial^2}{\partial
\tau^2}-2\frac{\tanh\tau}{r^2}\frac{\partial}{\partial \tau}+
\frac{1}{r^2\cosh^2\tau}\triangle_{\theta,\varphi},
\end{eqnarray}
where with $\triangle_{\theta,\varphi}$ is denoted the angular part
of the usual Laplace operator
\begin{eqnarray}
\label{angular}
\triangle_{\theta,\varphi}=\frac{1}{\sin\theta}\frac{\partial}{\partial\theta}
\left(\sin\theta\frac{\partial}{\partial\theta}\right)+\frac{1}{sin^2\theta}\frac{\partial^2}{\partial\varphi^2}.
\end{eqnarray}
The system of equations (\ref{syst}) in pseudo-spherical coordinates
becomes

\begin{eqnarray}
\label{systsp} \frac{3}{r}\frac{\partial E_z}{\partial
r}+\frac{\partial^2E_z}{\partial r^2}-
\frac{1}{r^2}\frac{\partial^2E_z}{\partial
\tau^2}-2\frac{\tanh\tau}{r^2}\frac{\partial E_z}{\partial \tau}+\nonumber\\
\frac{1}{r^2\cosh^2\tau}\triangle_{\theta,\varphi}E_z+
\gamma(|E_z|^2+|E_c|^2-|B_l|^2)E_z=0 \nonumber\\
\frac{3}{r}\frac{\partial E_c}{\partial
r}+\frac{\partial^2E_c}{\partial r^2}-
\frac{1}{r^2}\frac{\partial^2E_c}{\partial
\tau^2}-2\frac{\tanh\tau}{r^2}\frac{\partial E_c}{\partial \tau}+\\
\frac{1}{r^2\cosh^2\tau}\triangle_{\theta,\varphi}E_c+
\gamma(|E_z|^2+|E_c|^2-|B_l|^2)E_c=0\nonumber \\
\frac{3}{r}\frac{\partial B_l}{\partial
r}+\frac{\partial^2B_l}{\partial r^2}-
\frac{1}{r^2}\frac{\partial^2B_l}{\partial
\tau^2}-2\frac{\tanh\tau}{r^2}\frac{\partial B_l}{\partial \tau}+\nonumber \\
\frac{1}{r^2\cosh^2\tau}\triangle_{\theta,\varphi}B_l+
\gamma(|E_z|^2+|E_c|^2-|B_l|^2)B_l=0.\nonumber
\end{eqnarray}
Eqs. (\ref{systsp}) are solved using the method of separation of the
variables.

\begin{eqnarray}
\label{separat} E_i(r,\tau,\theta,\varphi)=
R(r)T_i(\tau)Y_i(\theta,\varphi)\\
B_l(r,\tau,\theta,\varphi)=
R(r)T_l(\tau)Y_l(\theta,\varphi),\nonumber
\end{eqnarray}
where $i=z,c$. We use an additional constrain on the angular and
 "spherical" time parts

\begin{eqnarray}
\label{contrains1}
|T_z|^2|Y_z|^2+|T_c|^2|Y_c|^2-|T_l|^2|Y_l|^2=const.
\end{eqnarray}
The condition (\ref{contrains1})  separates the variables. The
nonlinear terms appear in the radial part only. Thus the radial
parts obey the equation

\begin{eqnarray}
\label{RAD} \frac{3}{r}\frac{\partial R}{\partial
r}+\frac{\partial^2R}{\partial r^2}-\frac{A_i}{r^2}R+\gamma|R|^2R=0,
\end{eqnarray}
where $A_i,\; i=z,c,l$ are separation constants. We  look for
solutions which possess more clearly expressed localization than the
scalar soliton solution obtained in \cite{KOVACH1}

\begin{eqnarray}
\label{soliton2}   R=\frac{sech(ln(r^{\alpha}))}{r},
\end{eqnarray}
where $\alpha$, $\gamma$ and the separation constants $A_i,\; i =
z,c,l$ satisfy the relations $\alpha^2-1=A_i;\;2\alpha^2=\gamma.$
The corresponding $\tau$ - dependent part of the equations
(\ref{systsp}) are linear

\begin{eqnarray}
\label{tau}
\cosh^2\tau\frac{d^2T_i}{d\tau^2}+2\sinh\tau\cosh\tau\frac{dT_i}{d\tau}+\nonumber\\
(C_i-A_i\cosh^2\tau)T_i,=0,
\end{eqnarray}
where $i=z,c,l$ and $C_i$ are another separation constants connected
with the angular part  of the Laplace operator
$Y_i(\theta,\varphi)$. Only the following solutions of Eq.
(\ref{tau}) exist which satisfy the condition (\ref{contrains1}):
$T_z=\cosh\tau; \; T_c=\cosh\tau; \; T_l=\sinh\tau$, with separation
constants: for the electrical part $A_z=A_c=3; \; C_z=C_c=2$ and for
the magnetic part $A_l=3; \; C_l=0$. Thus the magnetic part of the
system of equations (\ref{systsp}) does not depend on the angular
components, i.e. $Y_l(\theta,\varphi)=0$, as for the  electrical
part $Y_z(\theta,\varphi),\; Y_c(\theta,\varphi)$ we have the
following linear system of equations

\begin{eqnarray} \label{YI}
\frac{\triangle_{\theta,\varphi} Y_i}{Y_i}=-2,
\end{eqnarray}
where now $i=z,c$. There are only two solutions of the Eq.
(\ref{YI}) which satisfy the condition (\ref{contrains1}):
$Y_z=\cos\theta;\; Y_c=\sin\theta\exp(i\varphi)$, Using the relation
between the separation constants $A_i$ and the real number $\alpha$
we obtain the following values for $\alpha$ and $\gamma$:
$\alpha^2=4;\; \alpha=\pm 2; \; \gamma=8$.

\begin{figure*}
\centering
\includegraphics[width=120 mm]{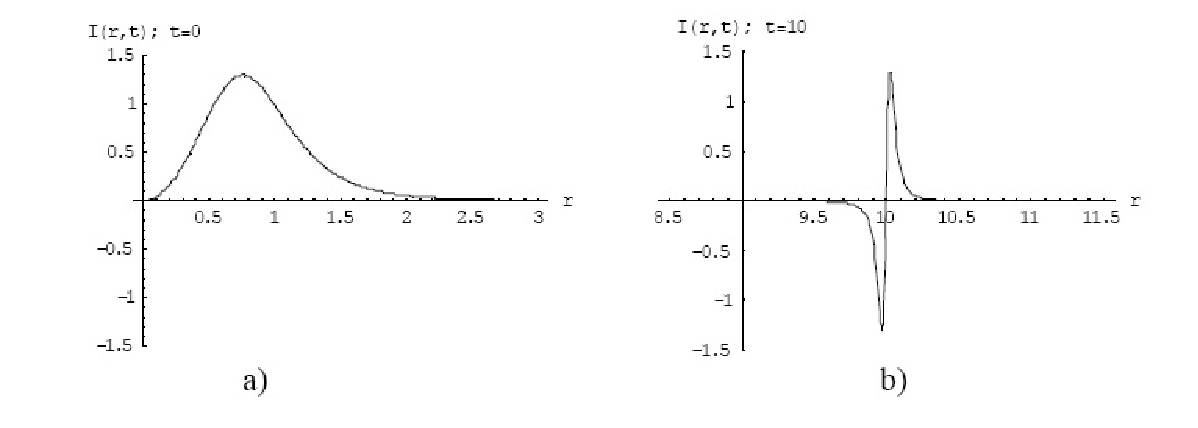}
\caption{Time evolution of the  intensity (\ref{INT1}) of the
solution (\ref{ELECTRON1}) of the nonlinear system of equations
(\ref{syst}) in Euler's vacuum ($c=1$) for $t=0$ and $t=10$
correspondingly. The nonlinear wave demonstrates entirely different
evolution, than the linear spherical one:  the shock wave preserves
its amplitude maximum and self-compresses  in $r$ direction.}
\end{figure*}

Finally, we can write the exact solution of the system of nonlinear
equations (\ref{syst}) which describes the propagation of a
electromagnetic wave in nonlinear vacuum

\begin{eqnarray}
\label{ELECTRON}
E_z(r,\tau,\theta)=\frac{sech(ln(r^{\pm 2}))}{r}\cosh\tau\cos\theta\nonumber\\
E_c(r,\tau,\theta,\varphi)=\frac{sech(ln(r^{\pm 2}))}{r}\cosh\tau\sin\theta\exp(i\varphi)\\
B_l(r,\tau)=\frac{sech(ln(r^{\pm 2}))}{r}\sinh\tau\nonumber.
\end{eqnarray}
If we rewrite the solution in Cartesian coordinates, it is not
difficult to show that the solution (\ref{ELECTRON}) of the system
(\ref{syst}) admits finite energy and the electrical part possesses
angular momentum $l=1$

\begin{eqnarray}
\label{ELECTRON1} E_z=\frac{2z}{r^4+1};\;
E_c=\frac{2(x+iy)}{r^4+1};\ B_l=\frac{2ict}{r^4+1},
\end{eqnarray}
where $r=\sqrt{x^2+y^2+z^2-c^2t^2}$. The  intensity profile of the
solution now becomes

\begin{eqnarray} \label{INT1}
I(x,y,z,t)= \frac{4(x^2+y^2+z^2-c^2t^2)}
{\left[\left(x^2+y^2+z^2-c^2t^2\right)^2+1\right]^2}.
\end{eqnarray}
For a comparison, in Fig. 1 we show the time evolution of the
intensity profile $I$ of a spherically symmetric analytical solution

\begin{eqnarray}
\label{ANZ4}
E\left(x,y,z,t\right)=1/\left[\frac{r^2}{r_0^2}+\left(1+\frac{ict}{r_0}\right)^2\right]
\end{eqnarray}
of the linear scalar wave equation

\begin{eqnarray}
\label {W} \Delta E= \frac{1}{c^2}\frac{\partial^2 E}{\partial t^2},
\end{eqnarray}
obtained recently by us applying the Fourier method. We have used
normalized scales $r_0=1$, $c=1$, times of evolution $t=0$ and
$t=10$. The initially localized amplitude function in the linear
case decreases with the generation of outside and inside fronts,
while the energy density distributes over the whole space in a
finite time. The evolution of the intensity profile (\ref{INT1}) is
presented in Figs. 2a and 2b for $t=0$ and $t=10$ correspondingly.
It is clearly seen from Fig. 2 that  solution (\ref{ELECTRON1})
describes a nonlinear shock wave in vacuum.  The wave admits
entirely different evolution than the linear spherical ones: as the
linear wave front enlarges spherically, the shock wave preserves its
amplitude maximum and self-compresses itself in $r$ direction.

In this paper the nonlinear vector wave equations in nonlinear
vacuum (\ref{EVAC1}) are solved through the method of separation of
the variables in a pseudo-spherical coordinate system. The obtained
analytical solution (\ref{ELECTRON1}) represents a spherical shock
wave with its own angular momentum $l=1$ for the electrical field.
Such high intensity wave can be generated not only from the laser
sources, but also in a nuclear reaction, where a nonlinear
polarization of virtual electron-positron pairs appears at the
beginning. If we compare the nonlinear vacuum shock wave with a
spherically symmetric solution of the linear wave equation, the
difference becomes obvious. While the spherically symmetric solution
of the linear wave equation forms inside and outside wave fronts and
the amplitude significantly decreases, the nonlinear shock wave
preserves the amplitude maximum and self-compress in $r$ direction.

\end{document}